\documentclass[traditabstract]{aa} 
\usepackage{graphicx}
\usepackage{amssymb, amsmath}
\usepackage{txfonts}
\usepackage{natbib}
\usepackage{lscape} 

\usepackage{txfonts}
\newcommand{\xmm}[0]{XMM-\textit{Newton}}
\newcommand{\cs}[0]{CoRoT-7}
\newcommand{\csb}[0]{CoRoT-7b}
\newcommand{\csd}[0]{CoRoT-7d}
\renewcommand{\csc}[0]{CoRoT-7c}
\newcommand{\Me}[0]{M$_{\oplus}$}
\newcommand{\ergcms}{erg$\,$cm$^{-2}$$\,$s$^{-1}$}
\newcommand{\ergs}{erg$\,$s$^{-1}$}

\begin{document}
   \title{The high-energy environment in the super-earth system CoRoT-7}

   \author{K. Poppenhaeger\inst{1}
          \and
          S. Czesla\inst{1}
          \and
          S. Schr\"oter\inst{1}
          \and
          S. Lalitha\inst{1}
          \and
          V. Kashyap\inst{2}
          \and
          J.H.M.M. Schmitt\inst{1}
          }

   \institute{Hamburger Sternwarte, Gojenbergsweg 112, 21029 Hamburg, Germany\\
              \email{katja.poppenhaeger@hs.uni-hamburg.de}   
   \and
   Harvard-Smithsonian Center for Astrophysics, 60 Garden Street, Cambridge, MA 02138, USA}

   \date{Received 23 November 2011; accepted 28 February 2012}

  \abstract
   {High-energy irradiation of exoplanets has been identified to be a key
   influence on the stability of these planets' atmospheres.
   So far, irradiation-driven mass-loss has been observed only in two Hot
   Jupiters, and
   the observational data remain even more sparse in the 
   super-earth regime.
   We present an investigation of the high-energy
   emission in the CoRoT-7 system, which hosts the first known
   transiting super-earth.
   To characterize the high-energy XUV radiation
   field into which the rocky planets CoRoT-7b and CoRoT-7c are immersed, we
   analyzed a $25$~ks \xmm \ observation of the host star.
   Our analysis yields the first clear ($3.5\sigma$) X-ray detection of CoRoT-7.
   We determine a coronal temperature of $\approx 3$~MK and an X-ray
   luminosity of $3\times 10^{28}$~erg$\,$s$^{-1}$. The level of XUV irradiation 
   on \csb \ amounts to $\approx 37\,000$~\ergcms. Current theories for planetary evaporation can only provide an order-of-magnitude estimate for the planetary mass loss; assuming that \csb \ has formed as a rocky planet, we estimate that \csb \ evaporates at a rate of about $1.3\times 10^{11}$~g$\,$s$^{-1}$ and has lost $\approx 4-10$ earth masses in total.
   }
   \maketitle
%

\section{Introduction}

After more than a decade of finding Jovian exoplanets,
planetary research has entered a new stage heralded by the
discovery of low-mass\,---\,possibly Earth-like\,---\,exoplanets.
\citet{Leger2009} reported on the detection of the first
known transiting super-earth, \csb. In quick succession, intense radial-velocity
follow-up revealed a second, \csc \ \citep{Queloz2009}, and potentially
a third planet, \csd \ \citep{Hatzes2010}, making
\cs \ a compact super-earth system.

\csb \ orbits its host star every $0.85$~d, has a mass of $\approx
7.4$~M$_{\oplus}$, and a semi-major axis of $0.017$~AU \citep{Hatzes2011}. \csc \ orbits at a
distance of $0.046$~AU, has roughly twice \csb's mass, and an
orbital period of $3.7$~d \citep{Queloz2009}.
The spectral properties of the host star, \cs, have been analyzed before
\citep{Leger2009, Bruntt2010}. \cs \ has an effective temperature of $5250 \pm
60$~K, a surface gravity of $\log(g) = 4.47 \pm 0.05$, and a metal overabundance
of \mbox{[M/H]$=0.12 \pm0.06$} with an abundance pattern consistent
with that of the Sun. \citet{Bruntt2010}
concluded that \csb \ orbits a main-sequence star of spectral type G8V-K0V.

The two super-earths \csb \ and \csc \ cause RV amplitudes of only
$\approx 5$~m$\,$s$^{-1}$.
Accurate RV measurements in the \cs \ system are made difficult because of the host star's activity; indeed, the analysis of \citet{Queloz2009} showed that the
dominating RV signal can be attributed to stellar rotation and, thus, activity.
Those authors found a mean $\log(R'(HK))$ index of $-4.612$ and a $2$\% variation
in phase with the $23$~d stellar rotation period.
This result is in line with
the optical variability observed in the CoRoT light-curve \citep{Leger2009}.
Pinpointing the exact mass of \csb \ has proven to be difficult because of the fairly high activity of the host star. The mass estimates of various authors differ by several Earth masses, see for example \citet{Queloz2009}, \citet{Hatzes2010}, and \citet{Ferraz-Mello2011}; here we use the mass determination from \citet{Hatzes2011} of $M_p = 7.4$~\Me, yielding a mean planetary density of $\rho_p = 10.4$~g\,cm$^{-3}$.

Interestingly, the known super-earths differ drastically in their
physical properties. For example, the Earth-like density of $7-10$~g\,cm$^{-3}$ found for \csb \ \citep{Bruntt2010, Hatzes2011} is in sharp
contrast to findings for GJ~1214b, which shows a density of only $1.9 \pm
0.4$~g\,cm$^{-3}$, suggesting the existence
a gaseous envelope \citep{Rogers2010, Nettelmann2011}. One possible source of
this differences is the
high-energy irradiation from the planets' host stars, which has been identified
as the main driver of planetary mass-loss and is therefore an important factor for
their evolution \citep[e.g.,][]{SanzForcada2010}.
Observational evidence for ongoing evaporation has been found for
two transiting Hot Jupiters: HD~209458b \citep{Vidal-Madjar2003} and HD~189733b
\citep{Lecavelier2010}. The accumulated effect of stellar high-energy emission
could have left fingerprints on today's planetary population
\citep{SanzForcada2010}. For \csb, \citet{Valencia2010} showed that
the accumulated planetary mass-loss may be as high as
$100$~M$_{\oplus}$ if the planet initially hosted a massive hydrogen-helium envelope and ca. $3-4$~M$_{\oplus}$ if the inital planetary density was similar to today's value.

To shed light on the possible atmospheric evaporation of \csb, we present here the first measurement of the high-energy emission of
the super-earth host-star \cs.

\section{Data analysis}

CoRoT-7 was observed with \xmm \ for approximately 25~ks on September 22,
2010 (ObsID~0652640201, PI V. Kashyap). The observation does not cover any
planetary transit. For our analysis of the X-ray data we used SAS version~11.0
and followed standard routines for data reduction. The observation, carried
out with the thin filter, is afflicted by strong background after
the first $10$~ks; the background contribution is particularly strong in the PN
detector. In our analysis we did not disregard time intervals of strong
background, as this would have meant losing most of the data, too. However, the EPIC background contributes strongest at energies above 2~keV and can therefore be reduced by choosing an approriate low-energy range where moderately active stars have their strongest X-ray emission.

\begin{figure}
\begin{center}
\vspace{0.5cm}
\includegraphics[width=0.35\textwidth]{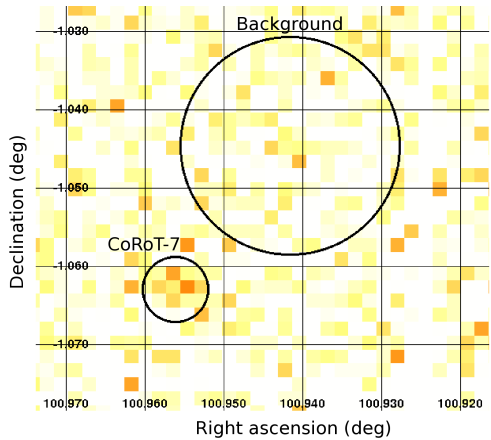}
\vspace{0.5cm}
\caption{Soft X-ray image of \cs, merged data from \xmm's
MOS1, MOS2, and PN camera in the 0.2-2~keV energy band.}
\label{Epicimage}
\end{center}
\end{figure}

In Fig.~\ref{Epicimage} we show the soft X-ray image of \cs. We extracted the source signal from a circular region with 15'' radius centered on
the nominal, proper-motion corrected position of \cs \ from SIMBAD; the background signal was measured in a source-free region close to \cs \ and subtracted from the signal in the source region. 

Our analysis was carried out in the soft energy band from $0.2-2$~keV, as
this is where stellar emission is typically strong for weakly to moderately
active stars as CoRoT-7, see for example \cite{Telleschi2005}. In this way, the strength of the background
in relation to the source signal was reduced.

\begin{figure}
\begin{center}
\includegraphics[width=0.48\textwidth]{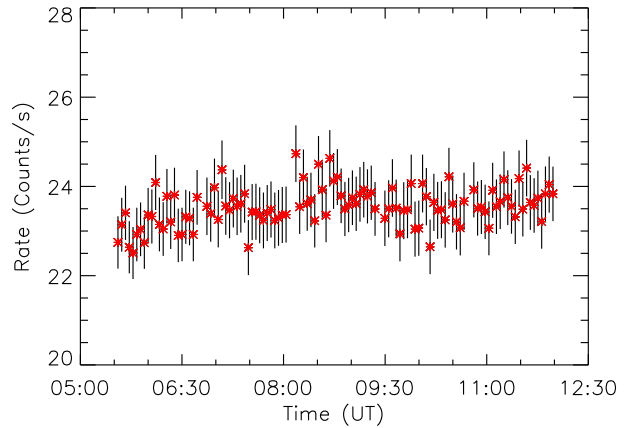}
\caption{UV light curve of CoRoT-7 recorded with \xmm's Optical Monitor and
rebinned to 200~s time resolution.}
\label{fig:OMLC}
\end{center}
\end{figure}

\xmm's Optical Monitor (OM) was used in the fast mode 
with the UVW1 filter inserted during the observation. The UVW1 filter
covers a wavelength range of ca. 240-360~nm. In Fig.~\ref{fig:OMLC}, we show the
OM light-curve with 200~seconds binning obtained from the raw data.

To obtain a distance estimate from interstellar absorption-features, we
analyzed 46 optical VLT-UVES spectra of \cs \ obtained during five nights
(Sept. 14 2008, Dec. 28 2009, Jan. 03 2010, Jan. 07 2010, Feb. 07 2010;
programs 081.C-0413(C) and 384.C-0820(A)).
The data were reduced using the ESO
UVES-pipeline in version~4.4.8 \citep{Ballester2000}. These data have been analyzed before to study possible emission and absorption features of the planetary atmosphere, resulting in a nondetection and an upper limit of $2 - 6\times 10^{-6} L_\ast$ for planetary emission in the \ion{Ca}{i}, \ion{Ca}{ii}, and \ion{NaD}{} lines \citep{GuentherEike2011}.

\section{Results}

To determine the X-ray luminosity of \cs \ and the resulting evaporation rates,
the distance to the system is crucial.
Therefore, we start our discussion by deriving new and complementary distance
estimates based on interstellar absorption and the Wilson-Bappu effect. 

\subsection{The distance to \cs}
\label{sec:distance}
\begin{figure}
\begin{center}
\includegraphics[width=0.48\textwidth]{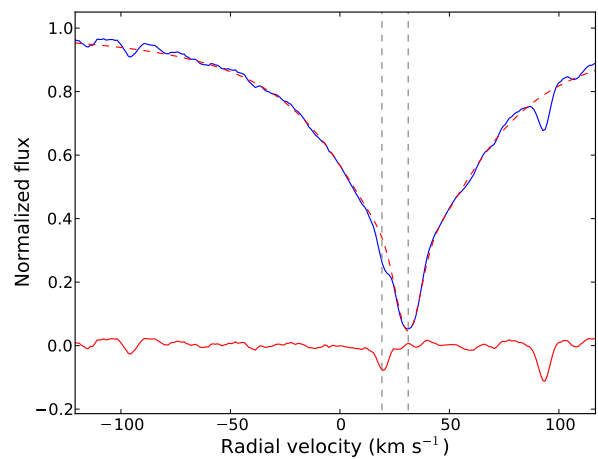}
\caption{Combined UVES spectra of \cs \ showing the \ion{Na}{i}~D$_2$~line
and interstellar absorption. After subtracting a
Voigt profile (dashed red) from the data (blue) the blueshifted absorption
feature becomes visible in the
residuals (solid red); the RV shift of
the interstellar feature ($+19$~km$\,$s$^{-1}$) and the \cs \ system
(+31~km$\,$s$^{-1}$) are marked by vertical, dashed lines. 
The absorption lines
in the outer parts of the line wing are due to
\ion{Fe}{i}.}
\label{NaDabs}
\end{center}
\end{figure}

In Fig.~\ref{NaDabs} we show an excerpt of the combined optical UVES spectra of \cs. In the blue wings of the \ion{Na}{i}~absorption-line doublet, we detected
an additional, narrow absorption feature, which shows a relative
displacement of $-12$~km\,s$^{-1}$ with respect to \cs \ in radial velocity. This corresponds to a barycentric velocity shift of
$+19$~km\,s$^{-1}$. Because the feature is visible in
all spectra and shows
no intrinsic variability, we attribute it to interstellar \ion{Na}{i}
absorption.


The equivalent widths (EWs) of the \ion{Na}~D$_2$ and \ion{Na}~D$_1$ features
are $12.1\pm1.6$~m\AA \ and $6.8\pm1.8$~m\AA. Using the line-ratio method
\citep{Stromgren1948}, we converted this measurement into a column density of
$7\times10^{10}$~cm$^{-2}$. \citet{Welsh2010} compiled a
catalog of interstellar \ion{Na}~absorption lines and provide a map
of the local interstellar medium, which we used to obtain a distance
estimate from the absorption measured along the line-of-sight.
Comparing the \ion{Na}~D$_{1,2}$ EWs with the
catalog data, we determined distance estimates of $160\pm80$~pc (D$_2$) and
$110\pm70$~pc (D$_1$) for \cs; the errors were estimated as the standard
deviation of the distances of stars with interstellar \ion{Na} EWs in a
$\pm5$~m\AA~band around the measured EWs
\citep[cf.,][]{Schroeter2011}.

Using the Wilson-Bappu effect, \citet{Bruntt2010} obtained an absolute visual
magnitude of $5.4\pm 0.6$~mag for \cs. Given the apparent brightness of
$V=11.7$~mag, this translates into a distance of $180\pm 50$~pc for
\cs.

Although the results derived from
\ion{Na}~D$_2$ and the photometric parallax favor a value of
$\approx 170$~pc, all estimates remain consistent with the distance of
$150\pm 20$~pc given by \citet{Leger2009}.
Therefore, we assume the same distance of $150$~pc in our analysis.

   \begin{table}
      \caption[]{X-ray net source counts of CoRoT-7 in the soft energy band (0.2-2 keV). Net counts obtained by subtracting the  background signal after scaling it to the same area (factor 0.09). 
      }
        \label{counts}
    \begin{center}
    \begin{tabular}{l r r r r}
    \hline\hline
			& MOS 1+2	& PN	& EPIC merged	\\ \hline
    Source region	& 183		& 567	& 750		\\
    Background		& 1389		& 5578	& 6967		\\
    Net counts		& 58		& 65	& 123	\\
    Significance	& $3.4\sigma$	& $2.1\sigma$ & $3.5\sigma$ \\ \hline

    \end{tabular}
    \end{center}
   \end{table}

\subsection{X-ray and UV properties}

We report a clear X-ray detection of CoRoT-7. The merged EPIC
image (MOS1, MOS2, and PN) shows an X-ray excess at the nominal position of
CoRoT-7 (see Fig.~\ref{Epicimage}).
The detailed source and background count numbers are given in
Table~\ref{counts}. From the merged data, we derived a detection significance of
$3.5\sigma$ in the soft energy band (0.2-2~keV).

Owing to the strong background signal a detailed analysis of the source's
temporal variability and spectral properties remained impossible.
To obtain an estimate of the coronal temperature despite this, we determined a
hardness ratio from the collected source counts in the MOS1, MOS2, and PN detector. Using a soft band of $0.2-0.7$~keV
and a hard band of $0.7-2.0$~keV, we determined a hardness ratio of
$HR = -0.22^{+0.25}_{-0.22}$; the given errors denote the equal-tail 68\% credibility interval.

\begin{figure}
\begin{center}
\includegraphics[width=0.48\textwidth]{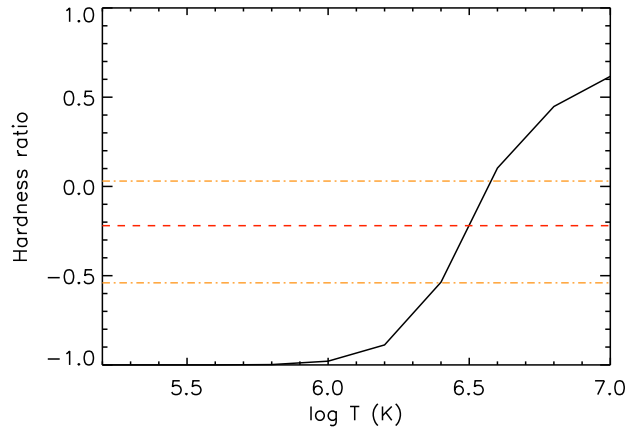}
\caption{Expected hardness ratio in \xmm's MOS and PN detectors, depending on
the mean coronal temperature (black); measured hardness ratio of CoRoT-7 (red
dashed) shown with 68\% credibility interval (orange dash-dotted).}
\label{hardness}
\end{center}
\end{figure}

We proceeded by calculating theoretical predictions for the hardness ratio
assuming different coronal temperatures. The modeling was carried out using
Xspec~v12.0 and the result is shown in Fig.~\ref{hardness} along with our
measurement. From this analysis, we can constrain the coronal temperature of CoRoT-7 to be
between $2.5$ and $3.7$~MK,
which is compatible with the coronal temperatures seen in moderately
active stars such as the Sun at its maximum activity. Consequently, we assume a coronal temperature of 3~MK to calculate CoRoT-7's
X-ray luminosity. 

At a stellar distance of 150~pc, interstellar hydrogen might produce noticeable absorption of the X-ray flux. According to \cite{Ferlet1985}, the interstellar \ion{Na} column density is a suitable tracer for hydrogen. Our measurement of the sodium column density $N(\ion{Na})=7\times 10^{10}$~cm$^{-2}$ converts to a hydrogen column density of \mbox{$N(\ion{H}{i}+\mathrm{H}_2)=3\times 10^{19}$~cm$^{-2}$}. Assuming a distance of 150~pc to CoRoT-7 (see Sect.~\ref{sec:distance}), this corresponds to a number density of about $0.1$~cm$^{-3}$ compatible with typical densities observed in the Local Cavity \citep{Cox2005}. 

Using WebPIMMS, we converted the mean count rate of a single \xmm \ MOS instrument of $1.16\times 10^{-3}$~counts\,s$^{-1}$ into an
unabsorbed X-ray flux of $1.0\times 10^{-14}$~ergs\,cm$^{-2}$\,s$^{-1}$ in the $0.2-2$~keV energy band, taking into account the encircled energy fraction of $68\%$ for our 15'' extraction region.
Given a
distance of 150~pc, this translates into an X-ray
luminosity of $3\times10^{28}$~erg\,s$^{-1}$. This is well in line with previous estimates derived from chromospheric \ion{Ca}{ii} H and K line emission, which yielded a predicted X-ray luminosity of $\approx 2\times10^{28}$~erg\,s$^{-1}$ \citep{Poppenhaeger2011AN}.
For a star of spectral type late G to early K,
CoRoT-7 displays a typical moderate level of coronal activity, which is also observed in stars with solar-like activity cycles \citep{HempelmannRobrade2006, Favata2008}:
with a bolometric
correction of $-0.437$ \citep{Flower1996}, we derived a bolometric luminosity of
\mbox{$\log L_{bol}=33.3$}
and, therefore, an X-ray activity indicator, $\log L_X/L_{bol}$, of $-4.8$.

The UV light-curve seen with the OM does not show strong flares (see
Fig.~\ref{fig:OMLC}). 
To determine the level and slope of the light curve, 
we fitted it with a linear model with free
offset and gradient and determined errors using a Markov-Chain Monte-Carlo
(MCMC) analysis. 
The mean count-rate amounts to
23.5(23.4-23.6)~counts\,s$^{-1}$, where the $95$\% credibility interval is
given in parenthesis. Using the count-to-energy conversion factors for a star of
spectral type K0 from the \xmm \ handbook, we calculated a flux of
$9\times 10^{-15}$~erg\,cm$^{-2}$\,s$^{-1}$ in the covered band of 240-360~nm.
Combining this with a distance of $150$~pc, we obtain a luminosity of $2.4\times
10^{28}$~erg\,s$^{-1}$ in the same band.

According to our analysis, the slope amounts to
\mbox{$8(1.8-14)\times 10^{-2}$~counts$\,$h$^{-1}$}, indicating a slight
increase in UV luminosity during the observation, which may be due to
intrinsic short-term variability induced by active regions or rotational modulation.
Although a temporal analysis of the X-ray flux was
not possible, the lack of pronounced UV variability indicates that it can be
interpreted as quiescent emission.

\section{Discussion: Evaporation of \csb \ and \csc}

Although an X-ray luminosity of $\approx 3\times 10^{28}$~\ergcms \ does not place \cs \
among the very active stars, the extreme proximity of its planets exposes them
to high levels of high-energy irradiation.

Given semi-major axes of $0.017$~AU and $0.046$~AU for \csb \ and \csc \
\citep{Queloz2009}, we calculated X-ray fluxes of $f_{X,b} = 3.7\times 10^{4}$~\ergcms \ and
$f_{X,c} = 5\times 10^{3}$~\ergcms \ at the orbital distances of the planets.  
Assuming a solar X-ray luminosity of $10^{27}$~\ergs, these fluxes exceed the
solar X-ray flux on Earth by factors of $100\,000$ and $14\,000$.

We now derive an order-of-magnitude estimate for the amount of mass that has been evaporated from \csb \ due to the stellar high-energy emission. Planetary evaporation is thought to be mainly driven by X-ray and EUV irradiation. The process is hydrodynamical \citep{Tian2005, Murray-Clay2009} and is much more efficient than pure Jeans escape. The approach has been refined for expanded absorption radii and mass-loss effects through Roche lobe overflow \citep{Lammer2003, Erkaev2007}. However, there are still considerable uncertainties about the dynamics of the process. In addition, we assume here that \csb \ has formed as a rocky planet and that the orbital distance is stable; there are other calculations that consider planetary migration as well as \csb \ being the solid core of an evaporated gaseous planet, which consequently derive very different mass loss histories \citep{Jackson2010}.  Here, we apply the same formula as used by \cite{SanzForcada2011} and \cite{Valencia2010}. The latter authors argue that the mass-loss rates, originally estimated for gaseous planets, can in principle remain true also for rocky planets, because the rate of sublimation in close vicinity to the star should be able to counterbalances the mass-loss:

\begin{equation}
\dot{M} = \frac{3 \epsilon F_{XUV}}{4 G \rho_p K},
\end{equation}
where $F_{XUV}$ is the incident XUV flux at the planetary orbit, $G$ is the gravitational constant, $\rho_p$ the mean planetary density, which we assume to be $10.4$~g\,cm$^{-3}$ according to \citet{Hatzes2011}, and $\epsilon$ is a factor to account for heating efficiency of the planetary atmosphere, with $\epsilon=1$ denoting that all incident energy is converted into particle escape. Several authors chose $\epsilon=0.4$ \citep{Valencia2010, Jackson2010}, inspired by observations of the evaporating Hot Jupiter HD~209458b, and we follow their approach. However, this can only be a rough estimate of the true heating efficiency, which cannot be stated more precisely without detailed models of the dynamics of the exoplanetary atmosphere. $K$ is a factor for taking into account effects from mass loss through Roche lobe overflow; we neglect these effects by assuming $K=1$. This is justified because the approximate Roche lobe radius \citep{Eggleton1983} is about three times today's radius of \csb \ . If the planetary mass had been three times as large initially, the Roche lobe radius would still have exceeded the initial planetary radius, as modeled by \citet{Seager2007} for rocky planets, by a factor of approximately two. Roche-lobe influenced mass loss was therefore likely insignificant for the planetary evolution. 

Because we have not measured \cs's EUV flux, we use a relation between X-ray ($0.1-2.5$~keV) and EUV
($100-920$~\AA) flux \citep{SanzForcada2011}:

\begin{equation}
\log L_{EUV} = (4.80 \pm 1.99) + (0.860 \pm 0.073) \log L_X.
\end{equation}

We measured the X-ray luminosity in the 0.2-2~keV band to be $3\times 10^{28}$~\ergs; we used WebPIMMS to extrapolate this to the required energy band of $0.1-2.5$~keV, the result being $3.3\times10^{28}$~\ergs. This leads to an EUV luminosity of $L_{EUV} = 2.1\times 10^{29}$~\ergs, and thus a combined XUV (X-ray and EUV) luminosity of $L_{XUV} = 2.4\times
10^{29}$~\ergs of \cs, which is about ten times higher than the X-ray luminosity alone.
Substituting this number into Eq.~1 yields an estimate of $1.3\times 10^{11}$~g$\,$s$^{-1}$ for \csb's current
mass-loss rate.

This number is higher than the one derived by
\citet{Valencia2010}, who estimated the XUV flux purely from an age-activity relation for solar-like stars given by \cite{Ribas2005}, since no X-ray detection of the star was available. \cs \ has a slightly sub-solar mass, being of spectral type G8-K0; however, for our order-of-magnitude estimate of the mass loss, the difference in activity evolution of G and early K stars is small \citep{Lammer2009} and can be neglected. \cite{Valencia2010} estimate the XUV luminosity as $5\times10^{28}$~\ergs. Assuming a lower planetary density than in this work, $4-8$\,g\,cm$^{-3}$, they derived a lower mass loss rate of $0.5-1\times 10^{11}$\,g\,s$^{-1}$.

For \csc, we estimate that the mass-loss rate should be about
an order of magnitude lower than for \csb \ due to the weaker irradiation. Because this planet does, however, not transit, its radius---and thus density---remains unknown.

Planetary mass-loss rates have been measured for two Hot Jupiters so far by using the transit depth in the hydrogen Ly~$\alpha$ line as an indicator for extended and escaping planetary atmospheres. \cite{Vidal-Madjar2003} reported a lower limit for the mass loss rate of HD~209458b of $\sim 10^{10}$\,g\,s$^{-1}$. Similarly, \cite{Lecavelier2010} derived a mass loss rate between $10^9$ and $10^{11}$\,g\,s$^{-1}$ for HD~189733b. The orders of magnitude are compatible with our calculations, even if these observational results were obtained for gaseous, not rocky planets.

Based on our mass-loss rate determination for \csb, we can derive an order-of-magnitude estimate for the total planetary mass-loss over time. 
The possible stellar age range for \cs \ is given by \cite{Leger2009} as $1.2-2.3$~Gyr; we adopt here an age of $1.5$~Gyr. Along the lines of \cite{Valencia2010} we assume a constant density of \csb. Likewise, for the stellar activity evolution, we assume that the combined X-ray and EUV luminosity increases at younger stellar ages $\tau$ by a factor of $(\tau/\tau_\ast)^{-1.23}$ with $\tau_\ast$ being the current stellar age in Gyr \citep{Ribas2005}, and that the activity remains at a constant level for ages younger than 0.1~Gyr. Inserting the time-variable XUV flux into Eq.~1 and integrating over 1.5~Gyr yields a total mass loss of ca. $6$~\Me. This is higher than the result of \cite{Valencia2010} of $3-4$~\Me, who used a lower stellar XUV flux and lower planetary density. If we consider the  uncertainty in the age determination of \cs, our estimated mass loss range extends to $4-10$~\Me.  Given the current uncertainties in the analytical models for planetary evaporation, we consider the possible range for the true mass loss of \csb \ to be even wider.

\section{Conclusion}

We report the first X-ray detection of \cs, host star to at
least two close-in super-earths. After veryfying the
distance estimate of $150$~pc of \cite{Leger2009}, we converted the X-ray flux
of $10^{-14}$~\ergcms \ measured with \xmm \ into a luminosity of
$3\times10^{28}$~\ergs (0.2-2~keV).

An X-ray activity indicator of $\log L_X/L_{bol}=-4.8$, combined with a likely coronal temperature
of $\approx 3$~MK, characterizes \cs \ as a moderately active star. 
Employing an analytical model for planetary evaporation --- with the caveat that many aspects of this evaporation are not yet understood ---, the combined X-ray and EUV irradiation can be converted into an order-of-magnitude estimate of the planetary mass loss. We derive a current mass loss rate of $1.3\times 10^{11}$~g$\,$s$^{-1}$ for the closest planet, \csb . Assuming that the planet was formed as a rock and integrating
the likely activity history of the \cs \ system, we estimate that \csb \ has 
suffered a total mass loss of $\approx 4-10$~Earth masses.

\begin{acknowledgements}
K.P. and S.L. acknowledge funding from the German Research Foundation (DFG) via Graduiertenkolleg 1351. S.C. and S.S. acknowledge financial support from DLR grant 50OR0703.
\end{acknowledgements}

\bibliographystyle{aa}
\bibliography{../katjasbib.bib}

\begin{thebibliography}{34}
\expandafter\ifx\csname natexlab\endcsname\relax\def\natexlab#1{#1}\fi

\bibitem[{{Ballester} {et~al.}(2000){Ballester}, {Modigliani}, {Boitquin},
  {Cristiani}, {Hanuschik}, {Kaufer}, \& {Wolf}}]{Ballester2000}
{Ballester}, P., {Modigliani}, A., {Boitquin}, O., {et~al.} 2000, The
  Messenger, 101, 31

\bibitem[{{Bruntt} {et~al.}(2010){Bruntt}, {Deleuil}, {Fridlund}, {Alonso},
  {Bouchy}, {Hatzes}, {Mayor}, {Moutou}, \& {Queloz}}]{Bruntt2010}
{Bruntt}, H., {Deleuil}, M., {Fridlund}, M., {et~al.} 2010, \aap, 519, A51

\bibitem[{{Cox}(2005)}]{Cox2005}
{Cox}, D.~P. 2005, \araa, 43, 337

\bibitem[{{Eggleton}(1983)}]{Eggleton1983}
{Eggleton}, P.~P. 1983, \apj, 268, 368

\bibitem[{{Erkaev} {et~al.}(2007){Erkaev}, {Kulikov}, {Lammer}, {Selsis},
  {Langmayr}, {Jaritz}, \& {Biernat}}]{Erkaev2007}
{Erkaev}, N.~V., {Kulikov}, Y.~N., {Lammer}, H., {et~al.} 2007, \aap, 472, 329

\bibitem[{{Favata} {et~al.}(2008){Favata}, {Micela}, {Orlando}, {Schmitt},
  {Sciortino}, \& {Hall}}]{Favata2008}
{Favata}, F., {Micela}, G., {Orlando}, S., {et~al.} 2008, \aap, 490, 1121

\bibitem[{{Ferlet} {et~al.}(1985){Ferlet}, {Vidal-Madjar}, \&
  {Gry}}]{Ferlet1985}
{Ferlet}, R., {Vidal-Madjar}, A., \& {Gry}, C. 1985, \apj, 298, 838

\bibitem[{{Ferraz-Mello} {et~al.}(2011){Ferraz-Mello}, {Tadeu Dos Santos},
  {Beaug{\'e}}, {Michtchenko}, \& {Rodr{\'{\i}}guez}}]{Ferraz-Mello2011}
{Ferraz-Mello}, S., {Tadeu Dos Santos}, M., {Beaug{\'e}}, C., {Michtchenko},
  T.~A., \& {Rodr{\'{\i}}guez}, A. 2011, \aap, 531, A161

\bibitem[{{Flower}(1996)}]{Flower1996}
{Flower}, P.~J. 1996, \apj, 469, 355

\bibitem[{{Guenther} {et~al.}(2011){Guenther}, {Cabrera}, {Erikson},
  {Fridlund}, {Lammer}, {Mura}, {Rauer}, {Schneider}, {Tulej}, {von Paris}, \&
  {Wurz}}]{GuentherEike2011}
{Guenther}, E.~W., {Cabrera}, J., {Erikson}, A., {et~al.} 2011, \aap, 525, A24

\bibitem[{{Hatzes} {et~al.}(2010){Hatzes}, {Dvorak}, {Wuchterl}, {Guterman},
  {Hartmann}, {Fridlund}, {Gandolfi}, {Guenther}, \&
  {P{\"a}tzold}}]{Hatzes2010}
{Hatzes}, A.~P., {Dvorak}, R., {Wuchterl}, G., {et~al.} 2010, \aap, 520, A93

\bibitem[{{Hatzes} {et~al.}(2011){Hatzes}, {Fridlund}, {Nachmani}, {Mazeh},
  {Valencia}, {H{\'e}brard}, {Carone}, {P{\"a}tzold}, {Udry}, {Bouchy},
  {Deleuil}, {Moutou}, {Barge}, {Bord{\'e}}, {Deeg}, {Tingley}, {Dvorak},
  {Gandolfi}, {Ferraz-Mello}, {Wuchterl}, {Guenther}, {Guillot}, {Rauer},
  {Erikson}, {Cabrera}, {Csizmadia}, {L{\'e}ger}, {Lammer}, {Weingrill},
  {Queloz}, {Alonso}, {Rouan}, \& {Schneider}}]{Hatzes2011}
{Hatzes}, A.~P., {Fridlund}, M., {Nachmani}, G., {et~al.} 2011, \apj, 743, 75

\bibitem[{{Hempelmann} {et~al.}(2006){Hempelmann}, {Robrade}, {Schmitt},
  {Favata}, {Baliunas}, \& {Hall}}]{HempelmannRobrade2006}
{Hempelmann}, A., {Robrade}, J., {Schmitt}, J.~H.~M.~M., {et~al.} 2006, \aap,
  460, 261

\bibitem[{{Jackson} {et~al.}(2010){Jackson}, {Miller}, {Barnes}, {Raymond},
  {Fortney}, \& {Greenberg}}]{Jackson2010}
{Jackson}, B., {Miller}, N., {Barnes}, R., {et~al.} 2010, \mnras, 407, 910

\bibitem[{{Lammer} {et~al.}(2009){Lammer}, {Odert}, {Leitzinger},
  {Khodachenko}, {Panchenko}, {Kulikov}, {Zhang}, {Lichtenegger}, {Erkaev},
  {Wuchterl}, {Micela}, {Penz}, {Biernat}, {Weingrill}, {Steller}, {Ottacher},
  {Hasiba}, \& {Hanslmeier}}]{Lammer2009}
{Lammer}, H., {Odert}, P., {Leitzinger}, M., {et~al.} 2009, \aap, 506, 399

\bibitem[{{Lammer} {et~al.}(2003){Lammer}, {Selsis}, {Ribas}, {Guinan},
  {Bauer}, \& {Weiss}}]{Lammer2003}
{Lammer}, H., {Selsis}, F., {Ribas}, I., {et~al.} 2003, \apjl, 598, L121

\bibitem[{{Lecavelier Des Etangs} {et~al.}(2010){Lecavelier Des Etangs},
  {Ehrenreich}, {Vidal-Madjar}, {Ballester}, {D{\'e}sert}, {Ferlet},
  {H{\'e}brard}, {Sing}, {Tchakoumegni}, \& {Udry}}]{Lecavelier2010}
{Lecavelier Des Etangs}, A., {Ehrenreich}, D., {Vidal-Madjar}, A., {et~al.}
  2010, \aap, 514, A72

\bibitem[{{L{\'e}ger} {et~al.}(2009){L{\'e}ger}, {Rouan}, {Schneider}, {Barge},
  {Fridlund}, {Samuel}, {Ollivier}, {Guenther}, {Deleuil}, {Deeg}, {Auvergne},
  {Alonso}, {Aigrain}, {Alapini}, {Almenara}, {Baglin}, {Barbieri}, {Bruntt},
  {Bord{\'e}}, {Bouchy}, {Cabrera}, {Catala}, {Carone}, {Carpano}, {Csizmadia},
  {Dvorak}, {Erikson}, {Ferraz-Mello}, {Foing}, {Fressin}, {Gandolfi},
  {Gillon}, {Gondoin}, {Grasset}, {Guillot}, {Hatzes}, {H{\'e}brard}, {Jorda},
  {Lammer}, {Llebaria}, {Loeillet}, {Mayor}, {Mazeh}, {Moutou}, {P{\"a}tzold},
  {Pont}, {Queloz}, {Rauer}, {Renner}, {Samadi}, {Shporer}, {Sotin}, {Tingley},
  {Wuchterl}, {Adda}, {Agogu}, {Appourchaux}, {Ballans}, {Baron}, {Beaufort},
  {Bellenger}, {Berlin}, {Bernardi}, {Blouin}, {Baudin}, {Bodin}, {Boisnard},
  {Boit}, {Bonneau}, {Borzeix}, {Briet}, {Buey}, {Butler}, {Cailleau},
  {Cautain}, {Chabaud}, {Chaintreuil}, {Chiavassa}, {Costes}, {Cuna Parrho},
  {de Oliveira Fialho}, {Decaudin}, {Defise}, {Djalal}, {Epstein}, {Exil},
  {Faur{\'e}}, {Fenouillet}, {Gaboriaud}, {Gallic}, {Gamet}, {Gavalda},
  {Grolleau}, {Gruneisen}, {Gueguen}, {Guis}, {Guivarc'h}, {Guterman},
  {Hallouard}, {Hasiba}, {Heuripeau}, {Huntzinger}, {Hustaix}, {Imad},
  {Imbert}, {Johlander}, {Jouret}, {Journoud}, {Karioty}, {Kerjean},
  {Lafaille}, {Lafond}, {Lam-Trong}, {Landiech}, {Lapeyrere}, {Larqu{\'e}},
  {Laudet}, {Lautier}, {Lecann}, {Lefevre}, {Leruyet}, {Levacher}, {Magnan},
  {Mazy}, {Mertens}, {Mesnager}, {Meunier}, {Michel}, {Monjoin}, {Naudet},
  {Nguyen-Kim}, {Orcesi}, {Ottacher}, {Perez}, {Peter}, {Plasson}, {Plesseria},
  {Pontet}, {Pradines}, {Quentin}, {Reynaud}, {Rolland}, {Rollenhagen},
  {Romagnan}, {Russ}, {Schmidt}, {Schwartz}, {Sebbag}, {Sedes}, {Smit},
  {Steller}, {Sunter}, {Surace}, {Tello}, {Tiph{\`e}ne}, {Toulouse}, {Ulmer},
  {Vandermarcq}, {Vergnault}, {Vuillemin}, \& {Zanatta}}]{Leger2009}
{L{\'e}ger}, A., {Rouan}, D., {Schneider}, J., {et~al.} 2009, \aap, 506, 287

\bibitem[{{Murray-Clay} {et~al.}(2009){Murray-Clay}, {Chiang}, \&
  {Murray}}]{Murray-Clay2009}
{Murray-Clay}, R.~A., {Chiang}, E.~I., \& {Murray}, N. 2009, \apj, 693, 23

\bibitem[{{Nettelmann} {et~al.}(2011){Nettelmann}, {Fortney}, {Kramm}, \&
  {Redmer}}]{Nettelmann2011}
{Nettelmann}, N., {Fortney}, J.~J., {Kramm}, U., \& {Redmer}, R. 2011, \apj,
  733, 2

\bibitem[{{Poppenhaeger} \& {Schmitt}(2011)}]{Poppenhaeger2011AN}
{Poppenhaeger}, K. \& {Schmitt}, J.~H.~M.~M. 2011, Astronomische Nachrichten,
  332, 1052

\bibitem[{{Queloz} {et~al.}(2009){Queloz}, {Bouchy}, {Moutou}, {Hatzes},
  {H{\'e}brard}, {Alonso}, {Auvergne}, {Baglin}, {Barbieri}, {Barge}, {Benz},
  {Bord{\'e}}, {Deeg}, {Deleuil}, {Dvorak}, {Erikson}, {Ferraz Mello},
  {Fridlund}, {Gandolfi}, {Gillon}, {Guenther}, {Guillot}, {Jorda}, {Hartmann},
  {Lammer}, {L{\'e}ger}, {Llebaria}, {Lovis}, {Magain}, {Mayor}, {Mazeh},
  {Ollivier}, {P{\"a}tzold}, {Pepe}, {Rauer}, {Rouan}, {Schneider},
  {Segransan}, {Udry}, \& {Wuchterl}}]{Queloz2009}
{Queloz}, D., {Bouchy}, F., {Moutou}, C., {et~al.} 2009, \aap, 506, 303

\bibitem[{{Ribas} {et~al.}(2005){Ribas}, {Guinan}, {G{\"u}del}, \&
  {Audard}}]{Ribas2005}
{Ribas}, I., {Guinan}, E.~F., {G{\"u}del}, M., \& {Audard}, M. 2005, \apj, 622,
  680

\bibitem[{{Rogers} \& {Seager}(2010)}]{Rogers2010}
{Rogers}, L.~A. \& {Seager}, S. 2010, \apj, 716, 1208

\bibitem[{{Sanz-Forcada} {et~al.}(2011){Sanz-Forcada}, {Micela}, {Ribas},
  {Pollock}, {Eiroa}, {Velasco}, {Solano}, \&
  {Garc{\'{\i}}a-{\'A}lvarez}}]{SanzForcada2011}
{Sanz-Forcada}, J., {Micela}, G., {Ribas}, I., {et~al.} 2011, \aap, 532, A6+

\bibitem[{{Sanz-Forcada} {et~al.}(2010){Sanz-Forcada}, {Ribas}, {Micela},
  {Pollock}, {Garc{\'{\i}}a-{\'A}lvarez}, {Solano}, \&
  {Eiroa}}]{SanzForcada2010}
{Sanz-Forcada}, J., {Ribas}, I., {Micela}, G., {et~al.} 2010, \aap, 511, L8

\bibitem[{{Schr{\"o}ter} {et~al.}(2011){Schr{\"o}ter}, {Czesla}, {Wolter},
  {M{\"u}ller}, {Huber}, \& {Schmitt}}]{Schroeter2011}
{Schr{\"o}ter}, S., {Czesla}, S., {Wolter}, U., {et~al.} 2011, \aap, 532, A3+

\bibitem[{{Seager} {et~al.}(2007){Seager}, {Kuchner}, {Hier-Majumder}, \&
  {Militzer}}]{Seager2007}
{Seager}, S., {Kuchner}, M., {Hier-Majumder}, C.~A., \& {Militzer}, B. 2007,
  \apj, 669, 1279

\bibitem[{{Str{\"o}mgren}(1948)}]{Stromgren1948}
{Str{\"o}mgren}, B. 1948, \apj, 108, 242

\bibitem[{{Telleschi} {et~al.}(2005){Telleschi}, {G{\"u}del}, {Briggs},
  {Audard}, {Ness}, \& {Skinner}}]{Telleschi2005}
{Telleschi}, A., {G{\"u}del}, M., {Briggs}, K., {et~al.} 2005, \apj, 622, 653

\bibitem[{{Tian} {et~al.}(2005){Tian}, {Toon}, {Pavlov}, \& {De
  Sterck}}]{Tian2005}
{Tian}, F., {Toon}, O.~B., {Pavlov}, A.~A., \& {De Sterck}, H. 2005, \apj, 621,
  1049

\bibitem[{{Valencia} {et~al.}(2010){Valencia}, {Ikoma}, {Guillot}, \&
  {Nettelmann}}]{Valencia2010}
{Valencia}, D., {Ikoma}, M., {Guillot}, T., \& {Nettelmann}, N. 2010, \aap,
  516, A20

\bibitem[{{Vidal-Madjar} {et~al.}(2003){Vidal-Madjar}, {Lecavelier des Etangs},
  {D{\'e}sert}, {Ballester}, {Ferlet}, {H{\'e}brard}, \&
  {Mayor}}]{Vidal-Madjar2003}
{Vidal-Madjar}, A., {Lecavelier des Etangs}, A., {D{\'e}sert}, J., {et~al.}
  2003, \nat, 422, 143

\bibitem[{{Welsh} {et~al.}(2010){Welsh}, {Lallement}, {Vergely}, \&
  {Raimond}}]{Welsh2010}
{Welsh}, B.~Y., {Lallement}, R., {Vergely}, J., \& {Raimond}, S. 2010, \aap,
  510, A54

\end{thebibliography}

\end{document}